\documentclass[aip,reprint]{revtex4-1}

\usepackage[utf8]{inputenc}

\usepackage{graphicx}

\usepackage{xcolor}

\usepackage{colortbl}

\begin{document}


\title{Collisionless Shock Acceleration of protons in a plasma slab produced in a gas jet by the collision of two laser-driven hydrodynamic shockwaves.}



\author{J.-R. Marquès}
\email[]{jean-raphael.marques@polytechnique.fr}
\affiliation{LULI, CNRS, École Polytechnique, CEA, Sorbonne Université, Institut Polytechnique de Paris, F-91128 Palaiseau Cedex, France}

\author{L. Lancia}
\affiliation{LULI, CNRS, École Polytechnique, CEA, Sorbonne Université, Institut Polytechnique de Paris, F-91128 Palaiseau Cedex, France}

\author{P. Loiseau}
\affiliation{CEA, DAM, DIF, F-91297 Arpajon Cedex, France}
\affiliation{Université Paris-Saclay, CEA, LMCE, 91680 Bruyères-le-Chatel, France}

\author{P. Forestier-Colleoni}
\affiliation{LULI, CNRS, École Polytechnique, CEA, Sorbonne Université, Institut Polytechnique de Paris, F-91128 Palaiseau Cedex, France}
\affiliation{now at Université Paris-Saclay, CEA, LIDYL, 91191 Gif-sur-Yvette, France}

\author{M. Tarisien}
\affiliation{CENBG, CNRS-IN2P3, Université de Bordeaux, 33175 Gradignan Cedex, France}

\author{E. Atukpor}
\affiliation{CENBG, CNRS-IN2P3, Université de Bordeaux, 33175 Gradignan Cedex, France}

\author{V. Bagnoud}
\affiliation{GSI Helmholtzzentrum für Schwerionenforschung GmbH, Planckstraße 1, 64291 Darmstadt, Germany}
\affiliation{University of Darmstadt, Schloßgartenstr. 764289 Darmstadt, Germany}

\author{C. Brabetz}
\affiliation{GSI Helmholtzzentrum für Schwerionenforschung GmbH, Planckstraße 1, 64291 Darmstadt, Germany}

\author{F. Consoli}
\affiliation{ENEA Fusion and Technologies for Nuclear Safety Department, C.R. Frascati, Via Enrico Fermi 45, Frascati, Rome, Italy}

\author{J. Domange}
\affiliation{CENBG, CNRS-IN2P3, Université de Bordeaux, 33175 Gradignan Cedex, France}

\author{F. Hannachi}
\affiliation{CENBG, CNRS-IN2P3, Université de Bordeaux, 33175 Gradignan Cedex, France}

\author{P. Nicolaï}
\affiliation{CELIA, Université de Bordeaux–CNRS–CEA, Talence 33405, France}

\author{M. Salvadori}
\affiliation{ENEA Fusion and Technologies for Nuclear Safety Department, C.R. Frascati, Via Enrico Fermi 45, Frascati, Rome, Italy}

\author{B. Zielbauer}
\affiliation{GSI Helmholtzzentrum für Schwerionenforschung GmbH, Planckstraße 1, 64291 Darmstadt, Germany}

\date{\today}

\begin{abstract}
We recently proposed a new technique of plasma tailoring by laser-driven hydrodynamic shockwaves generated on both sides of a gas jet [J.-R. Marquès \textit{et al.}, Phys. Plasmas \textbf{28}, 023103 (2021)]. In the continuation of this numerical work, we studied experimentally the influence of the tailoring on proton acceleration driven by a high-intensity picosecond-laser, in three cases: without tailoring, by tailoring only the entrance side of the ps-laser, or both sides of the gas jet. Without tailoring the acceleration is transverse to the laser axis, with a low-energy exponential spectrum, produced by Coulomb explosion. When the front side of the gas jet is tailored, a forward acceleration appears, that is significantly enhanced when both the front and back sides of the plasma are tailored. This forward acceleration produces higher energy protons, with a peaked spectrum, and is in good agreement with the mechanism of Collisionless Shock Acceleration (CSA). The spatio-temporal evolution of the plasma profile was characterized by optical shadowgraphy of a probe beam. The refraction and absorption of this beam was simulated by post-processing 3D hydrodynamic simulations of the plasma tailoring. Comparison with the experimental results allowed to estimate the thickness and near-critical density of the plasma slab produced by tailoring both sides of the gas jet. These parameters are in good agreement with those required for CSA.
\end{abstract}


\maketitle 

\section{Introduction}\label{Intro}

Collisionless shocks are ubiquitous in astrophysical environments \cite{Blanford,Jones} such as the Earth’s bow shock, solar flares, interplanetary traveling shocks, or supernova remnants (SNRs) \cite{Caprioli,Koyama}. They are believed to be responsible for non-thermal particles \cite{Adriani,Spitkovsky_2} and gamma ray bursts \cite{Spitkovsky_1}.
Using scaling laws \cite{Drake_2000}, the physics of collisionless magnetized shocks in SNRs has been investigated experimentally using laser-produced plasmas \cite{Courtois,Woolsey}. Their capability to accelerate particles has been demonstrated \cite{Fiuza_2020}.

The formation of a collisionless electrostatic shock requires the creation of a localized region of higher pressure within a plasma with electron temperature $T_e$ much larger than the ion temperature $T_i$. As this region of high pressure (defined as the downstream region) expands, it can drive a shock wave into the surrounding lower-pressure plasma (defined as the upstream region). The shock wave front can accelerate upstream ions by reflecting them to twice the shock velocity if the shock potential is larger than the kinetic energy of the incoming ions in the shock-rest frame.

Since they are efficient accelerators of particles, there has been a growing interest in exploring laser-driven shocks as compact particle accelerators \cite{Silva,Fiuza_2012,Fiuza_2013,Palmer_2011}. Energetic ion beams from compact laser-produced plasmas have potential applications in many fields of science and medicine, such as particle physics \cite{Bulanov_2005}, fast ignition of fusion targets \cite{Roth,Temporal_PPCF,Temporal_PoP}, material science \cite{Boody}, proton radiography \cite{Borghesi,Li,Romagnani}, radiotherapy \cite{Bulanov_2002,Bulanov_2002b,Malka,Karsch}, and isotope generation for medical applications \cite{Spencer,Lefebvre_2006}.

Several schemes for laser-driven ion acceleration have been proposed: Target Normal Sheath Acceleration \cite{Snavely,Wilks2001}, Radiation Pressure Acceleration \cite{Esirkepov,Robinson,Macchi_2009,Qiao}, Breakout Afterburner Acceleration \cite{Yin,Henig}, that use over-dense targets (solid density foils, liquid jets), or Magnetic Vortex Acceleration \cite{Bulanov_2005,Nakamura,Park} and Collisionless Shock Acceleration (CSA), that use near-critical density (NCD) targets (exploded foils, gas jets). The production, at high repetition rate, of high-energy ion beams with a narrow energy spread and a low divergence still remains a challenge, and CSA could potentially offer these properties. Moreover, NCD plasmas can be generated using gas jets, offering several advantages such as to avoid the constraints of target replacement and realignment between the consecutive shots, to avoid the target debris that usually spoil the surrounding optics, or to allow the production of pure proton beams (impurity free, using H$_2$ gas).

Proton acceleration by CSA in a hydrogen gas jet was first demonstrated \cite{Palmer_2011,Najmudin} using CO$_2$ lasers, the low critical density ($n_c \approx 10^{19}$ cm$^{-3}$) associated with their long wavelength ($\lambda_0$ = 10 $\mu$m) allowing to exploit regular-pressure, mm-scale gas jets. A proton beam of $\sim$MeV energy was produced with narrow energy spread ($\sigma \sim 4 \%$) and low normalized emittance ($<$ 8 nm.rad). Until today, the maximum proton energy produced by CSA in a gas jet \cite{Haberberger} is 20 MeV. It was obtained from a CO$_2$ laser composed of several picosecond (ps) pulses, the early pulses serving to ionize the gas jet and to steepen the plasma density profile on the front side of the target by the radiation pressure of the laser, so that the following pulses interacted with a step-like density profile. Despite this promising proton acceleration, the pulse train, inherent to the laser system, varied from shot to shot, making the interactions challenging to reproduce. Instabilities such as laser filamentation and hosing from the leading pulses result in variable density profiles, which in turn lead to fluctuations in the resultant ion beam. A solution to generate a steep plasma density profile in a more stable manner is to tailor the near-critical gas target by a hydrodynamic shockwave (HSW) excited before the arrival of the main ps-pulse. This HSW can be excited by a low energy laser prepulse that can be focused inside the gas jet \cite{Tresca}, or on a solid target placed at the entrance side of the gas jet \cite{Chen-Y}. In both cases it has been demonstrated that for long density profiles ($>$ 40 $\mu$m), ion beams with broadband energy spectrum were produced, while for shorter plasma lengths ($<$ 20 $\mu$m), quasimonoenergetic acceleration of protons was observed.

Experiments using more widespread solid state ps-lasers ($\lambda_0 \approx$ 1 $\mu$m, $n_c \approx 10^{21}$ cm$^{-3}$) have been performed using high-pressure narrow gas jets tailored by a HSW \cite{Chen-SN,Puyuelo}, or exploded $\mu$m-size solid foils \cite{Pak,Tochitsky}. Compared to the proof-of-principle experiments with CO$_2$ lasers, the number of accelerated protons was substantially higher ($\sim 10^4 \times$) thanks to the larger $n_c$ and vector potential of the laser field ($a_0 \sim$ 4 to 25, versus $\sim1$ on the CO$_2$ experiments). Proton beams with peaks at energies up to 5.5 MeV from tailored gas jets and 54 MeV from exploded foils were obtained.

Despite the large differences in terms of laser or target types, this set of experiments showed that the shape and maximum energy of the ion distributions strongly depend on the profile and peak value of the plasma density. Ion beams with peaked energy distribution and similar velocity of ion species with different charge-to-mass ratios, consistent with CSA, were observed only for plasmas with a steep density gradient at the laser entrance side, and of NCD. To obtain a narrow energy spread by CSA \cite{Silva,Fiuza_2012,Fiuza_2013} it is crucial to have a uniform shock velocity and ion reflection, which implies a uniform electron temperature profile, only achieved by a quick recirculation of the heated electrons due to the space-charge fields at the front and at the back of the target. Therefore, the plasma width $L$ should be limited, which for a moderate Mach number shocks ($M \gtrsim 1$), reads $L_{opt} \sim \lambda_0 (m_i/m_e)^{1/2}$. Considering a hydrogen plasma and $\lambda_0$ = 1 $\mu$m yields $L_{opt} \sim$ 40 $\mu$m, and therefore relatively sharp gradients on both sides of the target.

To generate such thin NCD plasmas, we recently proposed a new tailoring method \cite{Marques,Bonvalet} based on a narrow gas jet coupled with two parallel nanosecond (ns) heating-lasers that propagate perpendicular to the main ps-laser. These ns-beams are focused on both sides of the gas jet. The sudden and localized ionization and heating occuring at their focii generates two cylindrical HSWs. The expansion of each wave has two effects: i) the part that expands towards vacuum expels the wing of the gas jet, ii) while the part propagating towards the center of the jet compresses the plasma. The collision at the jet center of the two waves generates a thin plasma slab with steep gradients. Our hydrodynamic and ion Fokker–Planck simulations indicated that with present high-density gas jets\cite{Henares}, this method allows the production of a thin ($\gtrsim 10$ $\mu$m) plasma slab, with an adjustable density up to $2\times 10^{22}$ cm$^{-3}$ (20 $n_c$ for a 1 $\mu$m laser). This tailoring scheme can be implemented at high repetition rate, in a debris free environment, and with different types of gas. Such thin-NCD targets attracted a lot of attention recently not only for use in CSA or other ion acceleration mechanisms \cite{Bulanov_2010}, but also for brilliant gamma-ray and electron-positron pair production \cite{Liu,Zhu,Rosmej}.

In this paper we present experimental results on proton acceleration in a H$_2$ gas jet laser-tailored by the technique recently proposed in our previous numerical work \cite{Marques,Bonvalet}. We study the influence of plasma tailoring on the energy and angular distributions of the accelerated protons, for three cases: without tailoring, by tailoring only the entrance side of the ps-beam, or both sides of the gas jet. We observed the transition from a transverse (to the ps-beam direction) acceleration with a low energy broad spectrum produced by Coulomb explosion, to a forward acceleration with a higher energy peaked spectrum, in good agreement with CSA. The spatio-temporal evolution of the plasma profile was characterized with optical shadowgraphy of a probe laser beam. These shadowgraphy were simulated by a post-processing (ray-tracing) of three-dimensional hydrodynamic simulations of the plasma tailoring, and compared with the experimental observations to estimate the size and density of the resulting plasma slab.

\section{Experimental setup}\label{Exp}
The work was performed on two laser facilities, PHELIX at GSI (Germany) and PICO2000 at LULI (France), both delivering laser pulses at a fundamental wavelength $\lambda_0 \sim$ 1 $\mu$m. The plasma was generated from a narrow, high density, supersonic jet of hydrogen gas expelled from a 250 $\mu$m output diameter nozzle. The two parallel ns-beams for plasma tailoring were focused 300 $\mu$m from each sides of the jet center. The ps-laser beam driving the proton acceleration was propagating at 90$^\circ$ from the ns-beams and focused at the center of gas jet. This experimental arrangement is sketched in Ref. [\cite{Marques}](Fig. 1).
The output of the gas nozzle was placed 400 $\mu$m above the propagation plane of the ps and ns laser beams. The gas density profile in this plane was $n_{atom}(r)$ (cm$^{-3}$.bar$^{-1}$) $\sim  6.2 \times 10^{17} [e^{-(r/186)^2} + 0.63 e^{-(r/132)}]$. The backing pressure was adjustable up to 1000 bars allowing, without laser tailoring, to reach $n_e \sim  n_c$ (the critical plasma density, $n_c$(cm$^{-3}$) $\sim 1.11 \times 10^{21} \lambda_0^{-2}$($\mu$m)).

The ps-beam was focused to a Gaussian-like focal spot of 10 $\mu$m FWHM (in intensity). The temporal profile was Gaussian with a Full Width at Half Maximum (FWHM) of 0.5 ps at GSI and 1 ps at LULI, and a contrast on the ns scale (Amplified Stimulated Emission, ASE) of $10^{11}$ and $10^7$ respectively. The peak intensity was respectively $8 \times 10^{19}$ and $2\times 10^{19}$ W/cm$^2$, corresponding to a normalized vector potential of $a_0 \sim$ 8 and 4. The Rayleigh length of the ps-beam was $z_R \gtrsim$ 250 $\mu$m, of the order or greater than the width of the gas jet. The ns-beams had a square-like temporal profile of $\sim$ 1 ns duration. Each ns-beam was focused to a spot of FWHM $\sim$ 67 $\mu$m (GSI) and 42 $\mu$m (LULI) containing up to 15 J and 6 J respectively.

On both experiments the hydrodynamics evolution of the plasma was controlled by two-dimensional shadowgraphy and interferometry of a probe laser beam of 0.5 $\mu$m wavelength propagating at 90° from the ps-beam, parallel to the ns-beams. This probe beam was collimated, with a diameter covering the entire interaction region. At GSI it was generated by the leakage of a dielectric mirror in the ps-beam path, sent to a delay-line and frequency doubled, leading to a  jitter-free probe of 0.5 ps duration. Its transverse intensity (shadowgraphy) and phase (interferometry) profiles at the exit of the plasma was recorded on 16-bit CCD cameras.
The probe beam at LULI had a Gaussian temporal profile of FWHM $\sim$ 7 ns. The shadowgraphy and interferometry at a given time (snap-shot) were recorded on Gated Optical Intensifier (GOI) coupled to 16-bit CCD cameras, with an integration window of 100 ps. The shadowgraphy profile along the ps-beam axis was imaged on the entrance slit of a streak camera. The long duration of the probe beam allowed to follow the propagation of the shockwaves, from their creation to their collision, and thus to check and adjust the synchronization of HSWs collision with the arrival of the ps-beam.
The shadowgraphy and interferometry diagnostics used the same collecting lens, with an angular aperture of $\pm$ 3.5$^\circ$ at LULI and $\pm$ 2$^\circ$ at GSI, leading to a spatial resolution better than 10 $\mu$m.
The focal spots of the ns-beams were recorded on a 12-bit CCD, allowing before shot to adjust them on both sides of the gas jet, as well as control their positions, shapes and transmitted energies on shot. The position of the ps-beam was controlled before shot on a 8-bit CCD. In addition to these "side-view" diagnostics (shadowgraphy, interferometry, focal spots), the interaction region was also imaged from the "bottom" on a 12-bit CCD. The second harmonic light (0.5 $\mu$m) emitted by the ps-beam along its propagation in the plasma was recorded on the "side-view" and "bottom-view" diagnostics, allowing to identify the regions where the density gradients or the laser intensity are high, or where beam refraction or self-focusing could occur. 
The energy spectrum of the protons accelerated from the H$_2$ plasma were recorded on image plates (Fuji BAS-MS type) coupled to magnetic spectrometers positioned at 0$^\circ$, 30$^\circ$ and 70$^\circ$ from the ps-laser axis on the LULI experiment, and at 25$^\circ$, 37$^\circ$ and 53$^\circ$ on the GSI experiment.


\section{Interaction without plasma tailoring}\label{no tailoring}
\begin{figure}
	\includegraphics[width=\columnwidth]{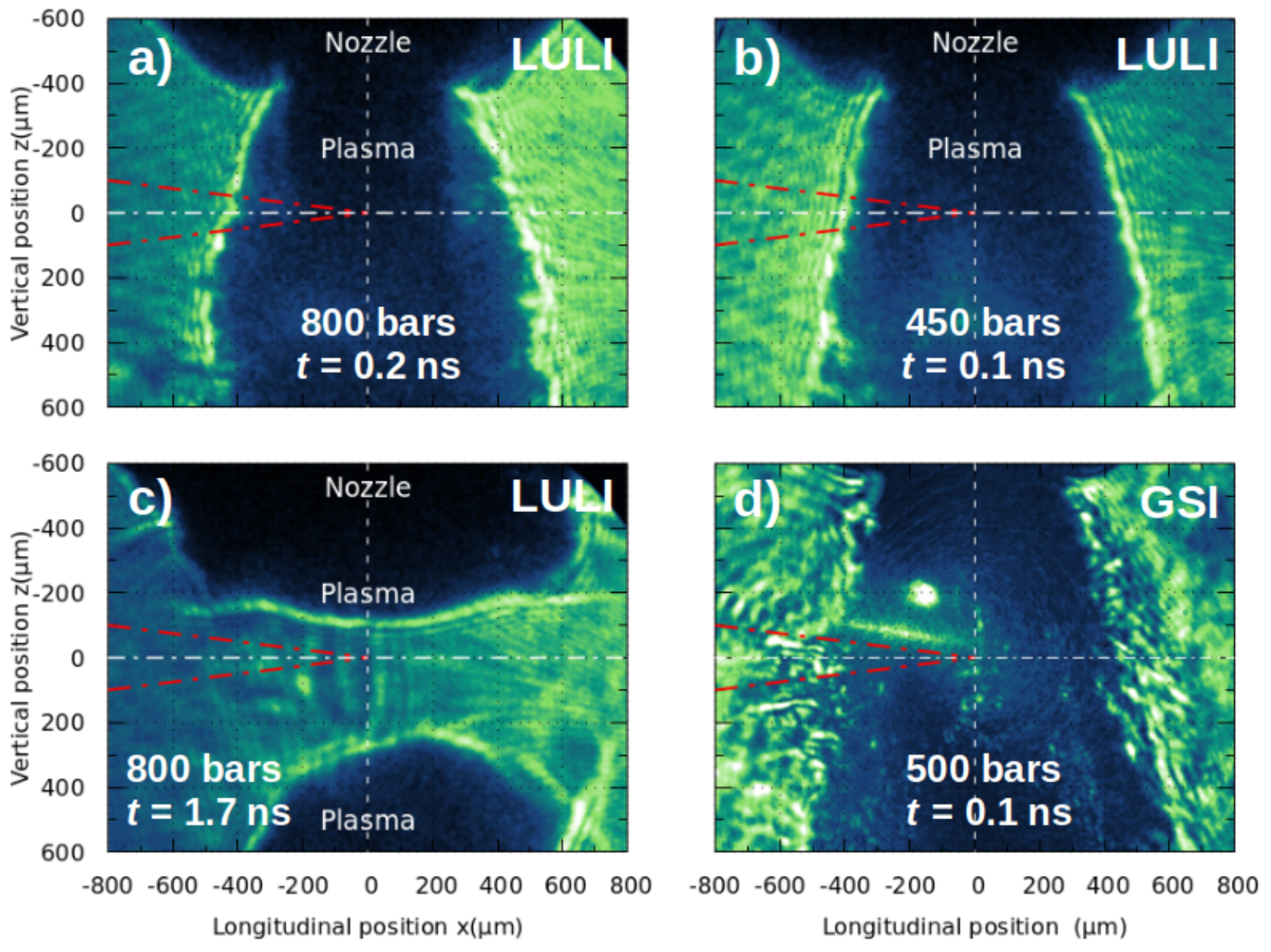}
	\caption{\label{ombro_LULI_GSI_sans_shaping}\textit{Probe beam shadowgraphies from the plasma generated by the interaction of the ps-beam with the high density H$_2$ gas jet, at different backing pressures, different times from the ps-pulse arrival, and for the LULI (a, b, c) or the GSI (d) ps-beam. On top of the images ($z<$ -400 $\mu$m) is the shadow of the nozzle. The ps-beam propagates at $z=0$ from left to right along the $x$-axis (horizontal white dashed line), and is focused (red dashed lines) at the center of the jet ($x$ = 0, $y$=0).}}
\end{figure}

Figure \ref{ombro_LULI_GSI_sans_shaping} shows two-dimensional (2D) space-resolved shadowgraphies (snapshot) of the probe beam from the plasma generated by the interaction of the ps-beam with the high density H$_2$ gas jet, at different backing pressures, times, and ps-beams (LULI and GSI). At 800 bars (Fig. \ref{ombro_LULI_GSI_sans_shaping} a) and c)) the expected peak density on the laser axis, assuming full ionization, is $n_e = 8 \times 10^{20}$ cm$^{-3}$ (0.8 $n_c$). Fig. \ref{ombro_LULI_GSI_sans_shaping}a) shows that the plasma at $t$ = 0.2 $\pm$ 0.1 ns after the ps-pulse arrival is much larger than the laser focal volume and extends on the whole gas jet volume, on a millimeter scale, along the laser axis ($x$) as well as transversely ($z$). Shadowgraphies of very similar size and shape are observed at 450 bars (Fig. \ref{ombro_LULI_GSI_sans_shaping} b)), on the LULI experiment, as well as on the GSI experiment (Fig. \ref{ombro_LULI_GSI_sans_shaping} d)) for which the probe pulse is shorter (0.5 ps) and the pump pulse has a $10^4$ times better ASE temporal contrast. The millimeter scale of these shadowgraphies is significantly larger than the FWHM of the gas jet, $\sim$ 270 $\mu$m at $z$ = 0, which will be discussed in section \ref{hydro_3D_simulations}. The prompt and localized energy deposition of the ps-pulse along its propagation leads to strong ponderomotive and thermal pressures which transversely expel the plasma from the laser axis, as observed in Fig. \ref{ombro_LULI_GSI_sans_shaping} c), 1.7 ns after the ps-pulse.

\begin{figure}
	\includegraphics[width=\columnwidth]{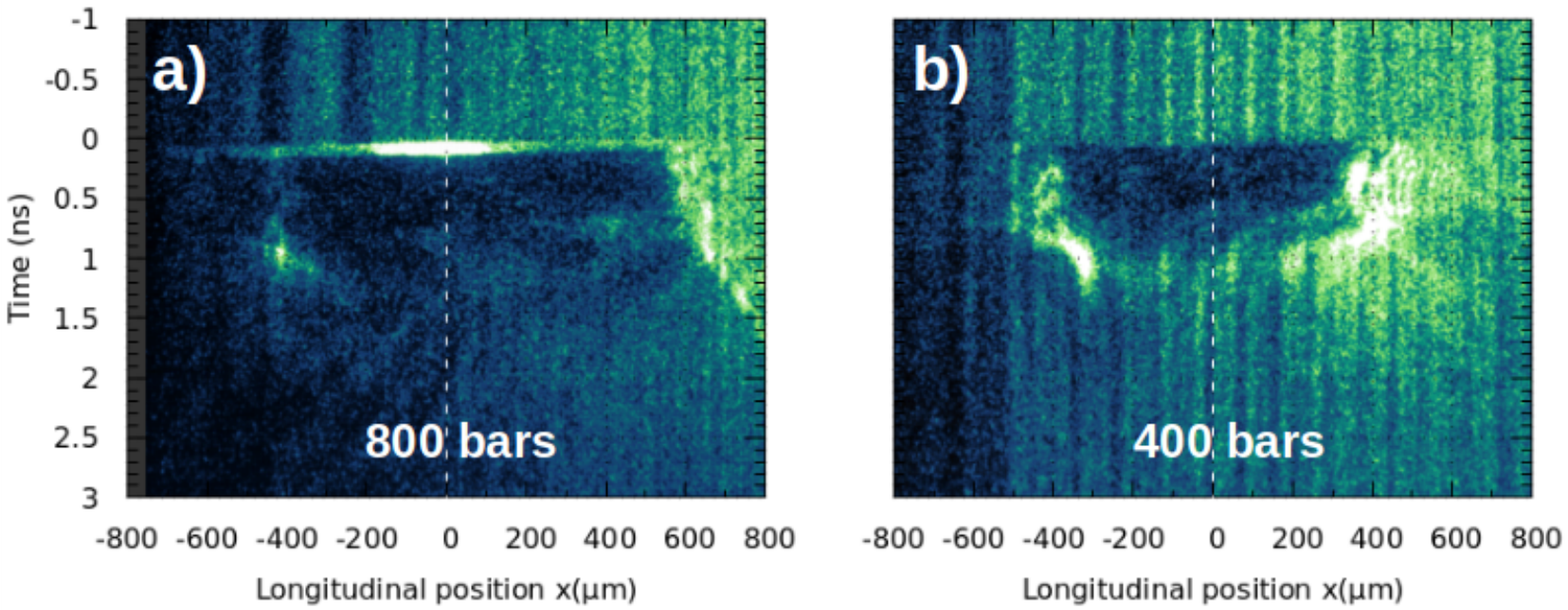}
	\caption{\label{ombro_streak_sans_shaping}\textit{Time-resolved (streak camera) shadowgraphies along the ps-beam axis ($x$, at $z=0$), for two gas jet pressures. The ps-pulse arrives at $t$ = 0 and propagates from left to right as in Fig. \ref{ombro_LULI_GSI_sans_shaping}.}}
\end{figure}

Time-resolved (streak camera) shadowgraphies along one dimension in space, the ps-beam axis ($x$, for $z=0$), are presented in Fig. \ref{ombro_streak_sans_shaping} for two gas jet pressures. Despite the factor two in pressure, these shadowgraphies are very similar. For $t$ $<$ 0 the probe beam is undisturbed, indicating that the ASE contrast is high enough to avoid ionization of the gas before the ps-pulse arrival. For 0 $<$ $t$ $<$ 1 ns, the shadowgraphy extends symmetrically from the focal position up to $x \sim \pm$ 400 $\mu$m, as observed on the snapshot in Fig \ref{ombro_LULI_GSI_sans_shaping}a) and b). At these edges, for the 800 bars case (Fig. \ref{ombro_streak_sans_shaping}-a)), the plasma density is $\sim 2 \times 10^{19}$ cm$^{-3}$ (0.02 $n_c$).
The bright line observed in Fig. \ref{ombro_streak_sans_shaping}a) at $t=0$, near the gas jet center (-200 $< x <$ +100 $\mu$m), is second harmonic emission produced by the ps-pulse, usually occurring in regions of strong laser intensity and/or density gradients. For 0 $< t <$ 1.5 ns, the longitudinal expansion of the right edge of the plasma ($x>500$ $\mu$m) can be observed. At $t \gtrsim 1$ ns the plasma expulsion induced by the ponderomotive and thermal pressures has lowered the density on the laser axis, and parts of probe beam start to be transmitted again. As in Fig. \ref{ombro_LULI_GSI_sans_shaping}c), at $t\sim1.7$ ns the plasma has been expelled from the laser axis and the probe beam is not disturbed anymore. 

\begin{figure}
	\includegraphics[width=\columnwidth]{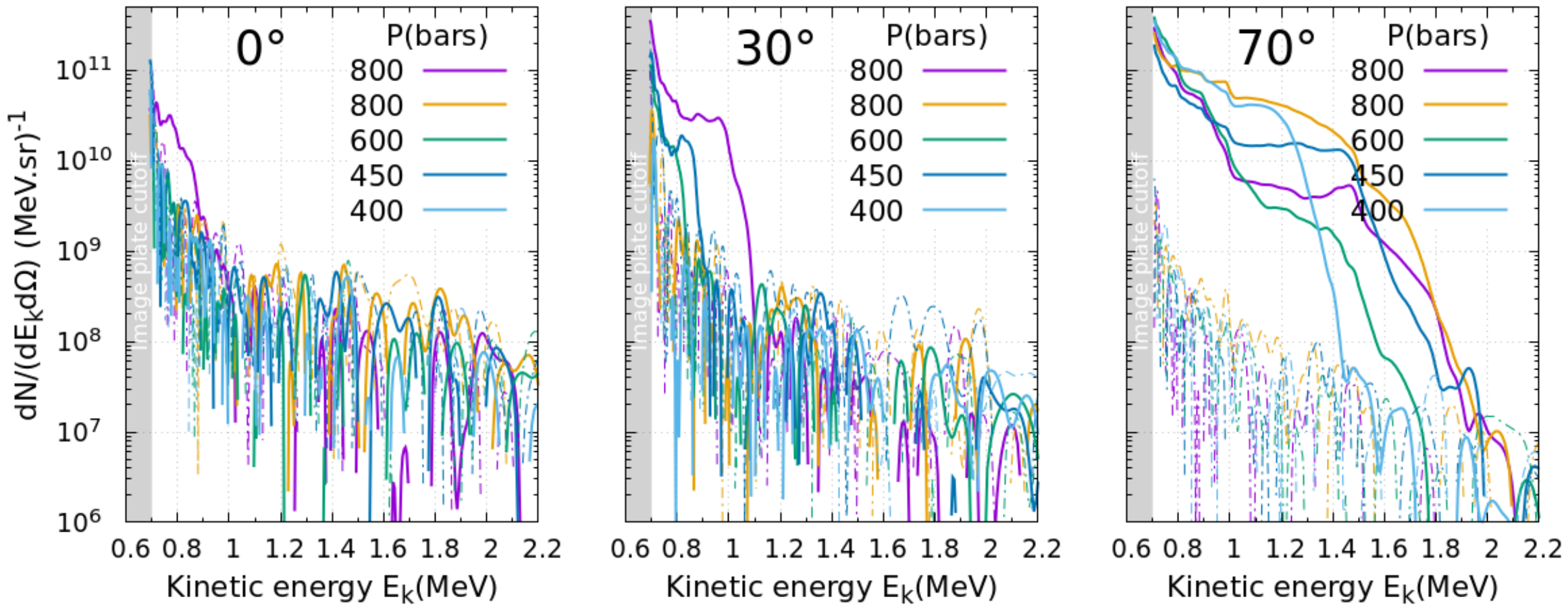}
	\caption{\label{Protons_wo_shaping}\textit{Proton spectra from the LULI experiment, without plasma tailoring, measured at 0$^{\circ}$, 30$^{\circ}$, and 70$^{\circ}$ from the ps-beam axis, and for different gas jet backing pressures. The dashed lines are the noise level of each spectrum. Same laser-plasma parameters as in Fig. \ref{ombro_LULI_GSI_sans_shaping} and \ref{ombro_streak_sans_shaping}.}}
\end{figure}

Proton spectra from the LULI experiment, measured at 0$^{\circ}$, 30$^{\circ}$, and 70$^{\circ}$ from the laser axis, are presented in Fig. \ref{Protons_wo_shaping} for the same laser-plasma parameters as Fig. \ref{ombro_LULI_GSI_sans_shaping} and \ref{ombro_streak_sans_shaping}. By directly focusing the laser on the gas jet, without plasma tailoring (no ns-beams), the acceleration along the laser axis is very weak (if not null), is larger at 30$^{\circ}$, and much more efficient at 70$^{\circ}$, almost perpendicularly to the laser axis. The spectrum has two components: an exponential part at low energy, followed by a plateau, extending up to 2 MeV at 70$^{\circ}$.Such an energy distribution could be the result of "Coulomb explosion": ions are accelerated by electrostatic forces caused by charge separation induced by the laser ponderomotive pressure in the underdense part of the plasma \cite{Sarkisov,Krushelnick}. The maximum energy that can be gained is the relativistic ponderomotive energy $U_p = m_e c^2 (\gamma-1)$, where $\gamma = (1+a_0^2 /2)^{1/2}$ is the relativistic factor of the electron quiver motion in the laser field. For the LULI experiment $a_0 \sim 4$, leading to $U_p \sim 1$ MeV, too low to explain the high energy part of the spectra. In addition, the ponderomotive energy depends only on the laser intensity, while the maximum energies observed at 0$^{\circ}$, 30$^{\circ}$, and 70$^{\circ}$ seem to increase with the plasma density. The plateau structure, the maximum energy value and its increase with the plasma density could be the signature of an acceleration by multiple collisionless shocks formed at high density \cite{Wei,Silva}.

\section{Interaction with a front face tailored plasma}\label{1beam-tailoring}
\begin{figure}
	\includegraphics[width=\columnwidth]{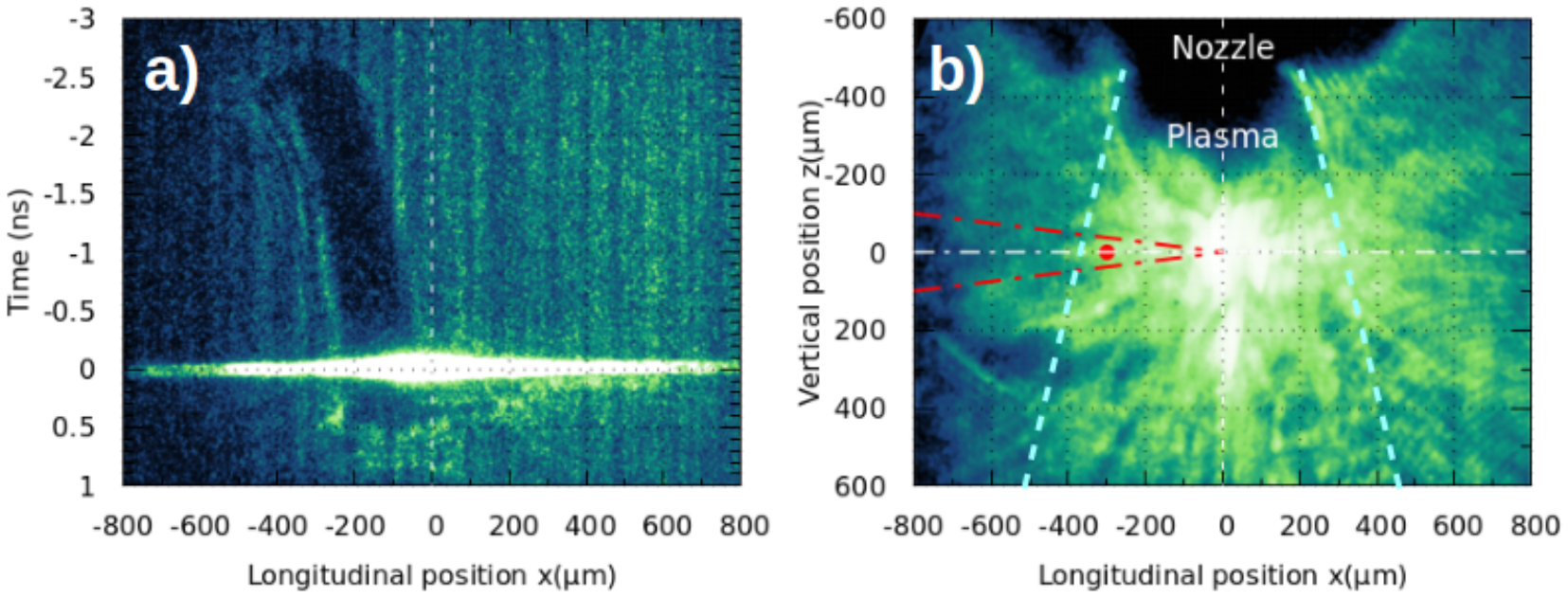}
	\caption{\label{ombro_LULI_shaping_face_avant}\textit{a) Time-resolved along the ps-beam axis (streak camera) and b) 2D space-resolved (GOI snapshot) shadowgraphy of the plasma tailored by one ns-beam, at the entrance side of the ps-beam, for a gas jet backing pressure of 200 bars. The ns-pulse propagates along the $y$ axis (perpendicular to the image plane) and focuses at $x=$ -300 $\mu$m, $z=$ 0 (red circle). It arrives in the gas jet $\sim$ 2.7 ns before the ps-pulse. The ps-beam propagates at $z=0$ from left to right along the $x$-axis (horizontal white dashed line in b)), and is focused (red dashed lines in b)) at the center of the jet ($x$ = 0, $y$=0). The time integration window of the snapshot in b) was $\sim$ 120 ps, centered on the ps-pulse arrival.}}
\end{figure}

The 1D time-resolved (streak camera) and 2D space-resolved (GOI snapshot) shadowgraphy of the plasma tailored by one ns-beam, at the entrance side of the ps-beam, is presented in Fig. \ref{ombro_LULI_shaping_face_avant}-a) and -b) respectively, for a gas pressure of 200 bars. The ns-pulse was sent in the gas jet $\sim$ 2.7 ns before the ps-pulse. It was propagating along the $y$ axis (perpendicular to the image plane) and focused at $x=$ -300 $\mu$m, $z=$ 0 (red circle in Fig. \ref{ombro_LULI_shaping_face_avant}-b)). The hydrodynamic evolution (on the ps-beam axis) of the plasma can be followed in Fig. \ref{ombro_LULI_shaping_face_avant}-a). For $t<$ -2.5 ns the probe beam is fully transmitted, indicating no pre-plasma. At $t\sim$ -2.7 ns the ns-beam promptly ionizes and heats the edge of the gas jet. A hydrodynamic shockwave is generated, which starts to push the plasma out of the ns-beam axis. At $-2.7 < t < -2.3$ ns one can observe the plasma motion towards vacuum ($x<$ -300 $\mu$m) and towards the jet center ($x>$ -300 $\mu$m). At later times, the plasma moving towards vacuum becomes too teneous to induce shadowgraphy, while the HSW propagating towards the jet center stays dense and can easily be followed. In contrast with the ps-pulse interaction that ionizes very quickly the whole gas jet volume (Fig. \ref{ombro_LULI_GSI_sans_shaping}), the ionization by the ns-beam concerns only its focal region, the rest of the jet staying in the gaseous state until the arrival of the HSW (the right side of the jet stays undisturbed).
The shockwave reaches the jet center at $t=$ 0. At that time the ps-pulse arrives, crosses the plasma created by the HSW, and then interacts with the high-density non-ionized part of the gas jet (at $x=$ 0), which induces strong second harmonic emission that saturates the camera. The 2D plasma extension at that time can be observed on Fig. \ref{ombro_LULI_shaping_face_avant}-b). The focal spot position of the ns-beam is indicated by the red circle. The time integration window of the GOI was $\sim$ 120 ps, centered on the ps-pulse arrival, so that the snapshot shows the plasma created by both the ns-beam and the ps-beam, as well as the strong second harmonic emission of the ps-pulse. Partly hidden by this flash, a croissant-like shape can be distinguished, which is the plasma generated by the expansion of the HSW. This initially cylindrical wave produces a circular shadowgraphy that has evolved to a croissant-like shape because $\sim$ half of the HSW expelled the plasma towards vacuum (left on the image), while the part propagating towards the center of the jet ionized and compressed this high density region. The horizontal extension of the croissant is larger on the bottom part of the image ($z>$ 0) because of the conical shape of the gas jet (larger far from the nozzle exit). The blue dashed-lines in Fig. \ref{ombro_LULI_shaping_face_avant}-b) show the edges of the shadowgraphy obtained for the case without tailoring (ps-beam only, Fig. \ref{ombro_LULI_GSI_sans_shaping}-a), -b), -d)). On the ns-beam side the plasma edge has been pushed towards the jet center.


\begin{figure}
	\includegraphics[width=\columnwidth]{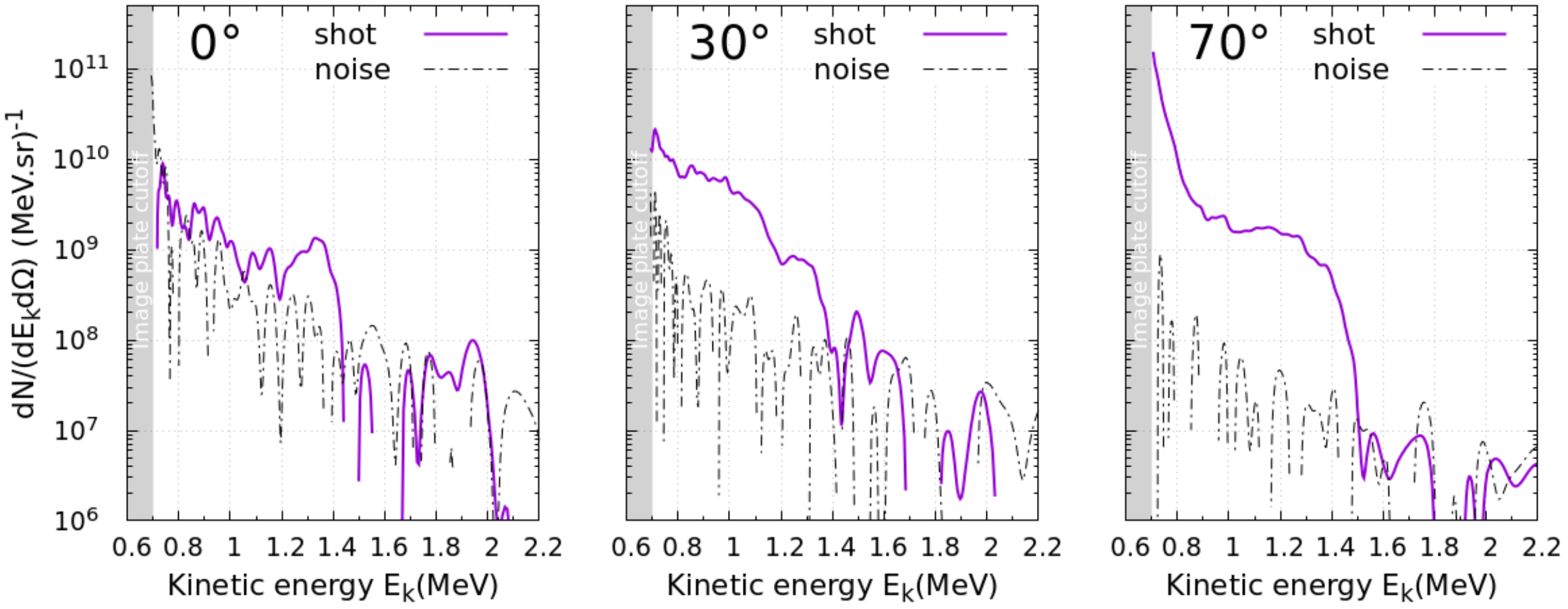}
	\caption{\label{Protons_shaping_face_avant}\textit{Proton spectra from the LULI experiment, associated with the shadowgraphy of Fig. \ref{ombro_LULI_shaping_face_avant}, for a plasma tailored at the entrance side of the ps-beam.}}
\end{figure}

Proton spectra associated with the shadowgraphy of Fig. \ref{ombro_LULI_shaping_face_avant} are presented in Fig. \ref{Protons_shaping_face_avant}. Tailoring the gas jet on the input side of the ps-beam improved the forward acceleration. The maximum proton energy at 0$^{\circ}$ and 30$^{\circ}$ both increased, while the transverse (70$^{\circ}$) acceleration became less efficient. In addition to the thermal exponential, the spectrum at 0$^{\circ}$ (Fig. \ref{Protons_wo_shaping}) has a peaked component centered near 1.35 MeV. This indicates the appearance of another acceleration mechanism: Collisionless Shock Acceleration. No acceleration (proton) was measured without the ps-beam.

\section{Interaction with a front and back face tailored plasma}\label{2beam-tailoring}

\begin{figure}
	\includegraphics[width=\columnwidth]{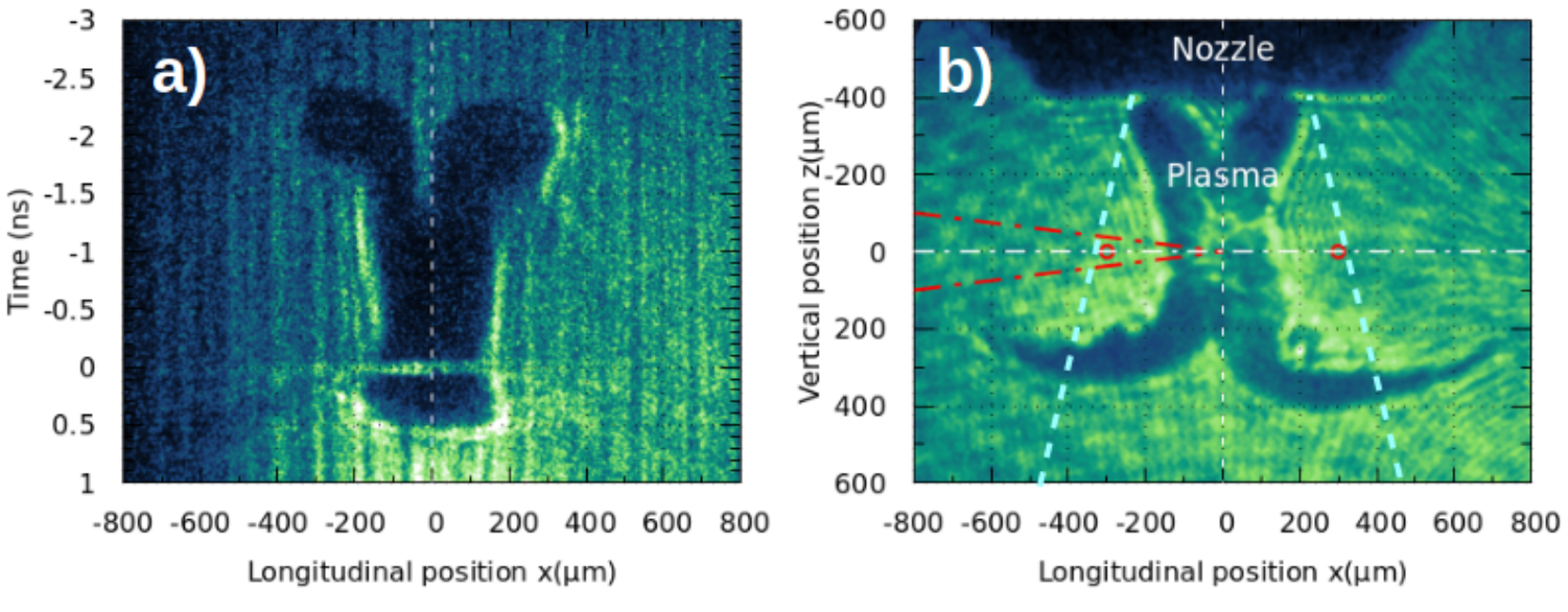}
	\caption{\label{ombro_LULI_shaping_deux_faces}\textit{a) Time-resolved along the ps-beam axis (streak camera) and b) 2D space-resolved (GOI snapshot) shadowgraphy of the plasma tailored on the two opposite sides, for a gas jet backing pressure of 200 bars. The energy in each ns-beam is 4.5 J. The time integration window of the snapshot in b) was $\sim$ 120 ps, ending just before the ps-pulse arrival. The blue dashed-lines indicate the edges of the shadowgraphy measured without plasma tailoring (Fig. \ref{ombro_LULI_GSI_sans_shaping}).}}
\end{figure}

\begin{figure}
	\includegraphics[width=\columnwidth]{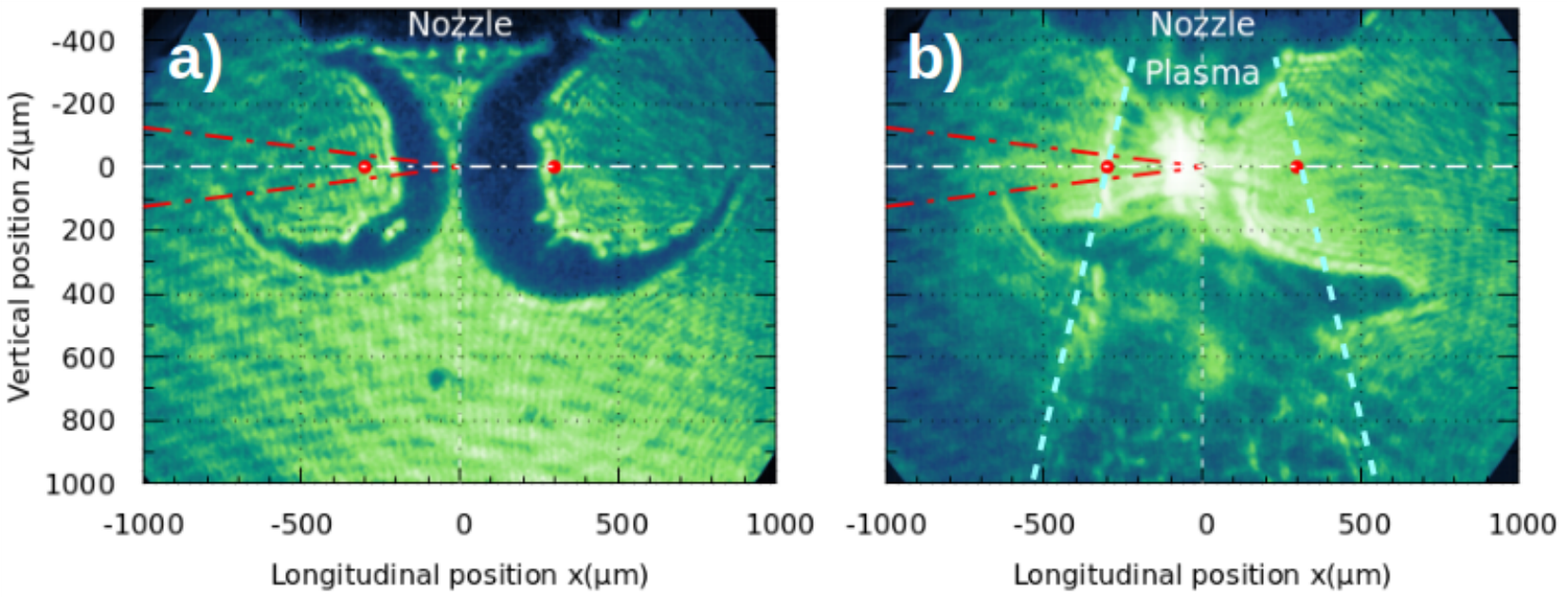}
	\caption{\label{ombro_LULI_shaping_deux_faces_deux_temps}\textit{2D space-resolved (GOI snapshot) shadowgraphies of the plasma tailored on the two opposite sides, for a gas jet backing pressure of 300 bars, at two different times: a) before the collision of the HSWs, and b) just after their collision and the arrival of the ps-pulse. The blue dashed-lines indicate the edges of the shadowgraphy measured without plasma tailoring (Fig. \ref{ombro_LULI_GSI_sans_shaping}).}}
\end{figure}

An example of shadowgraphies of the plasma tailored on the two opposite sides is presented in Fig. \ref{ombro_LULI_shaping_deux_faces}. The ns-beams arrive 2.5 ns before the ps-pulse. At this time we can see in Fig. \ref{ombro_LULI_shaping_deux_faces}-a) the prompt generation of the plasmas on both sides of the jet center, at the focal positions of the ns-beams ($x=\pm$ 300 $\mu$m), followed by the propagation of the HSWs. Only the dense parts converging towards the jet center are visible, the parts expanding towards vacuum are too teneous to generate a significant shadowgraphy. The central part of the jet stays in the gaseous state at least until $t\sim$ -1.4 ns, when the shadowgraphies of each HSW start to overlap. The waves collide at the jet center at $t\sim$ -100 ps, leading to a shadowgraphy with a minimum width of $\lesssim$ 300 $\mu$m, slightly before the ps-pulse arrival ($t=0$). This width is much smaller than the one observed in Fig. \ref{ombro_streak_sans_shaping} without plasma tailoring ($\sim$ 800 $\mu$m). As observed in Fig. \ref{ombro_streak_sans_shaping}-a) and \ref{Protons_shaping_face_avant}-a), the ps-pulse generates second harmonic light at its arrival in the plasma. After that time, the collision of the HSWs has ended and the plasma starts to expand. At $t\sim$ 0.6 ns the strong ponderomotive and thermal pressures driven by the ps-pulse have expelled the plasma from the laser axis, and the probe beam is transmitted again. The 2D space-resolved shadowgraphy of the HSWs just before the ps-beam arrival is presented in Fig \ref{ombro_LULI_shaping_deux_faces}-b). As already observed in Fig. \ref{ombro_LULI_shaping_face_avant}-b), the initially cylindrical expansion of each HSW from the axis of its driving ns-beam (red circles) has led to a croissant-like plasma of high density. The blue dashed-lines in the figure indicate the edges of the shadowgraphy measured without plasma tailoring (Fig. \ref{ombro_LULI_GSI_sans_shaping}). It shows that tailoring the gas jet from the two opposite sides allowed to i) significantly reduce the plasma density at the edge of the gas jet and, ii) generate a high-density narrow ($<$ 300 $\mu$m) plasma slab along the ps-beam axis. This is also illustrated in Fig. \ref{ombro_LULI_shaping_deux_faces_deux_temps} at a slightly higher pressure (300 bars), at two different times: before the collision of the HSWs (Fig. \ref{ombro_LULI_shaping_deux_faces_deux_temps}-a), and just after their collision and the arrival of the ps-pulse (Fig. \ref{ombro_LULI_shaping_deux_faces_deux_temps}-b). Before the ps-pulse arrival the plasma is visible only at the location of the HSWs, while after its arrival the region of the gas jet where the HSWs did not travel (bottom of the image, $z >$ 400 $\mu$m) is promptly ionized, leading in this region to a broad shadowgraphy comparable to the one observed without plasma tailoring  (Fig. \ref{ombro_LULI_GSI_sans_shaping}). On the opposite, around the ns-beam axis, where the plasma has been expelled by the HSWs, the ps-pulse does not create any additional perturbation of the shadowgraphy, indicating that it propagates almost freely, up to the HSWs colliding region where it encounters a high-density plasma slab and generates second harmonic emission (near $x\sim$ -100 $\mu$m).

\begin{figure}
	\includegraphics[width=\columnwidth]{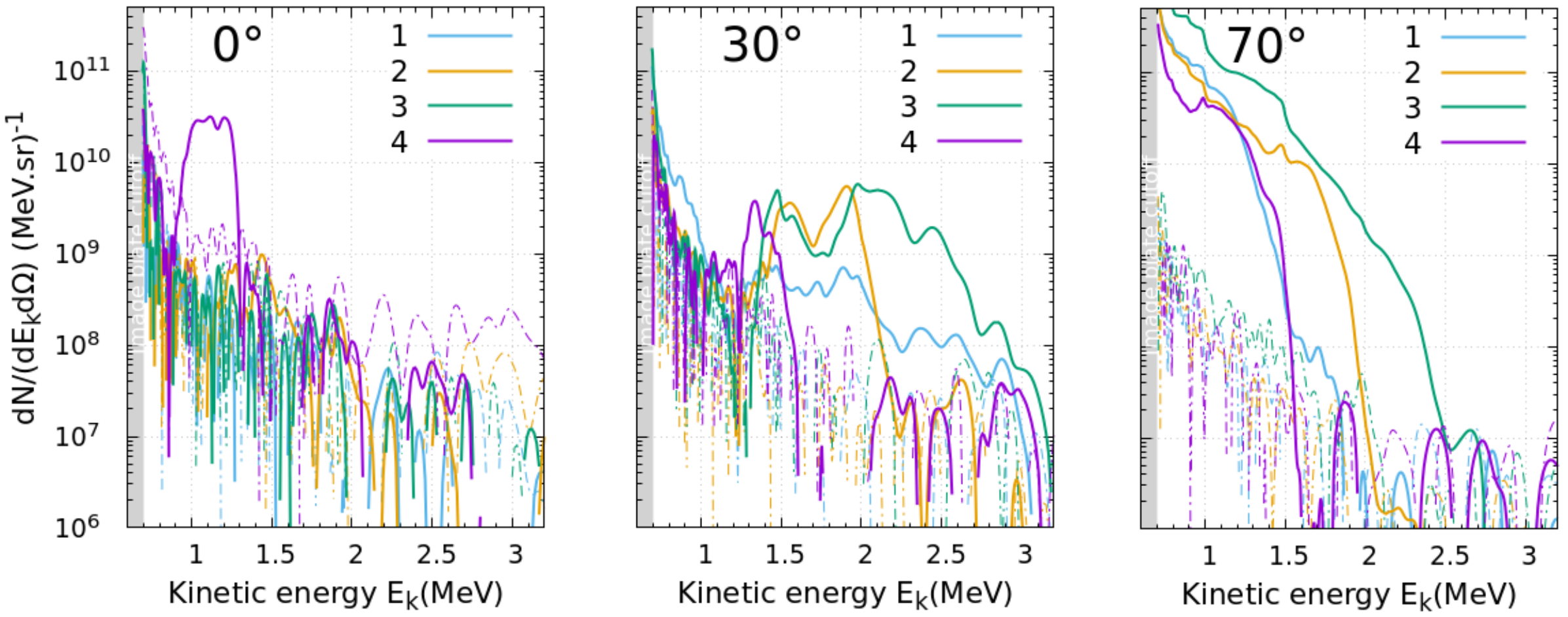}
	\caption{\label{Protons_shaping_deux_faces}\textit{Proton spectra from 4 laser shots on the LULI experiment, in the case of a plasma tailored on both sides (as in Fig. \ref{ombro_LULI_shaping_deux_faces}), and a gas jet backing pressure of 300 bars on shot 1 and 200 bars on shots 2 to 4. The dashed lines are the noise level of each spectrum.}}
\end{figure}

Examples of proton spectra obtained by tailoring the gas jet from the two opposite sides are presented in Fig. \ref{Protons_shaping_deux_faces}. The maximum energy and the shape of the spectra at 70$^{\circ}$ is quite similar to the ones obtained without tailoring (Fig. \ref{Protons_wo_shaping}). However, despite the shot-to-shot fluctuations, the tendency observed when tailoring only the entrance side (Fig. \ref{Protons_shaping_face_avant}) is enhanced when tailoring the two sides: the acceleration in the forward direction is improved and the energy distribution becomes more peaked. At 30$^{\circ}$ the portion of the spectra induced by CSA moved towards higher energies and is now clearly separated from the low energy thermal part. The maximum energy is more than doubled, reaching 3.2 MeV.

The number of protons accelerated above 1 MeV is also significantly increased at 0$^{\circ}$ and 30$^{\circ}$. The comparison with the single HSW case (Fig. \ref{Protons_shaping_face_avant}) demonstrates that the acceleration improvement is not only the result of a better propagation of the ps-pulse at the plasma entrance. Compared to the case without tailoring at 800 bars (Fig. \ref{Protons_wo_shaping}), the number of protons accelerated at 70$^{\circ}$ is similar or larger despite a lower backing pressure, 200-300 bars. This indicates that the collision of the HSWs has allowed to compensate the initially lower plasma density (the expected total compression factor \cite{Marques} is $\gtrsim$ 8), and enabled to reach a value at least equivalent to the one at 800 bars, close to $n_c$. It not only increases the number of protons available in the plasma, but helps to increase the ps-pulse absorption \cite{Wilks1992,Tsung}, thus the velocity of the collisionless shockwave and the maximum proton energies. Let us note that the lower backing pressure used for the tailoring cases was chosen to avoid the ns-beam refraction at the jet entrance and a better plasma expulsion by the HSW, while preserving the high density at the jet center thanks to the collision of the HSWs \cite{Marques}.

\begin{figure}
	\includegraphics[width=\columnwidth]{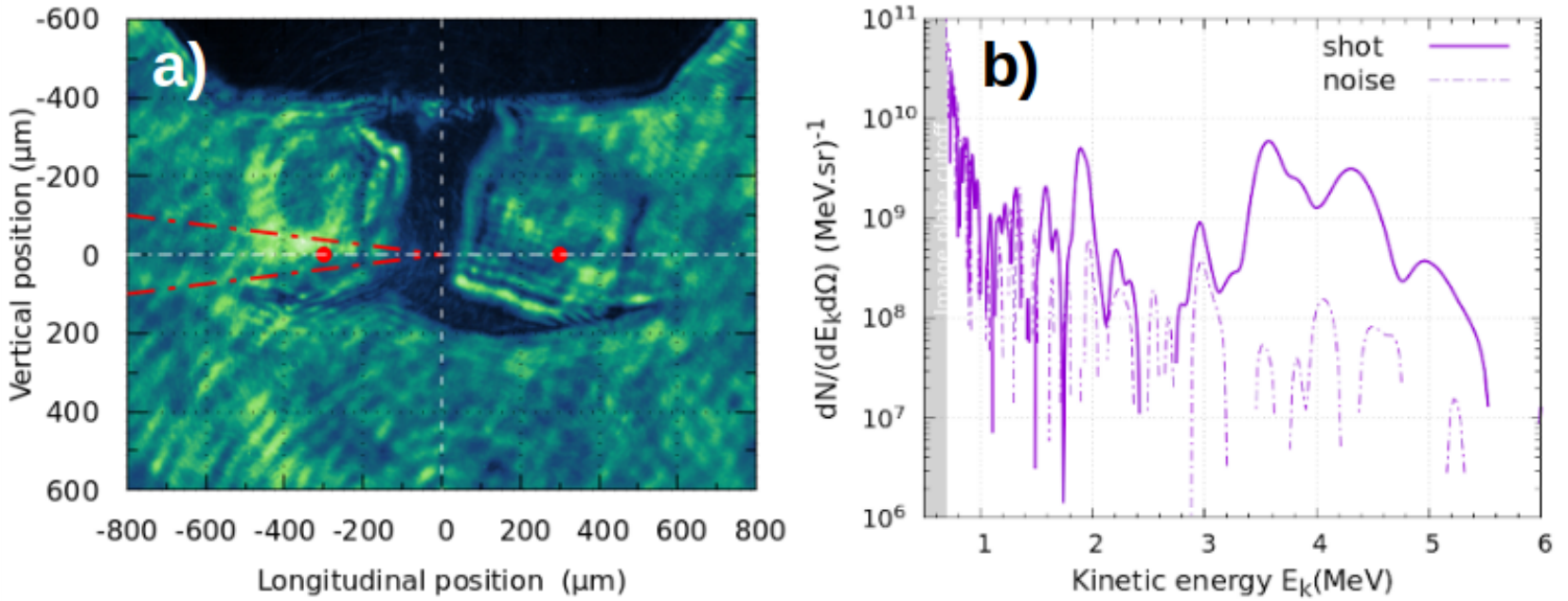}
	\caption{\label{ombro_spectres_GSI}\textit{Typical result of the GSI experiment: a) 2D space-resolved (GOI snapshot) shadowgraphy of the plasma tailored on both sides, b) associated proton spectrum measured at 50$^{\circ}$. The gas jet backing pressure was 100 bars. The shadowgraphy is 40 ps before the ps-pulse arrival.}}
\end{figure}

Very similar results were obtained on the GSI experiment. A 2D shadowgraphy at the collision of the two HSWs is presented in Fig. \ref{ombro_spectres_GSI}-a). The croissant-like shape of the HSWs is more square. This is due to the square shape of the wings of the ns-beam focal spots. Despite this difference, the sharpness and the width ($\sim$ 200 $\mu$m) of the plasma slab are very similar to the LULI experiment. The proton spectrum measured at 53$^{\circ}$ and associated to the interaction of the ps-pulse with this plasma slab is presented in Fig. \ref{ombro_spectres_GSI}-b). Like on the LULI experiment, the spectrum had a low energy exponential distribution without tailoring ($<$ 2 MeV), evolving to a peaked distribution of higher energy, up to 5.5 MeV, when two-sided tailoring is applied. Compared to the LULI experiment, the laser intensity was a factor $\sim$ 4 larger on GSI ($a_0 \sim$ 8 instead of 4), leading as expected to a factor $\sim$ 2 increase on the maximum energy of the protons.

Table \ref{Table_proton_numbers} summarizes the number of accelerated protons and the energy of the proton beam for the LULI spectra of Fig. \ref{Protons_wo_shaping}, \ref{Protons_shaping_face_avant} and \ref{Protons_shaping_deux_faces} associated with the three tailoring scenario. Despite the relatively strong shot-to-shot fluctuations, it demonstrates that tailoring the plasma from the two opposite sides improves the acceleration towards the laser direction, in terms of number of protons as well as of energy of the proton beam. The optimum acceleration in the forward direction also corresponds to the minimum transverse acceleration (last line in Table \ref{Table_proton_numbers}). Only one shot is presented for the one-side case. Nevertheless, several shots with the one-side tailoring were performed on the GSI experiment. They confirm the evolution of the acceleration observed on the LULI experiment between zero, one, or two sides tailoring. The shot-to-shot fluctuations observed on both experiments result mainly from variations on the plasma profile at the ps-pulse arrival, which can significantly affect the quality (uniformity, angle, Mach number) of the collisionless shockwave. These variations originate mainly from fluctuations on the dynamics of the HSWs which are sensitive to i) the energy in the wing of the ns-beam focal spot (see following section) and ii) the exact position of the focal spot on the edge of the gas jet density profile. For example, from the expression of $n_{atom}(r)$ given in section \ref{Exp}, the density is a factor 3.6 different between $r$ = 250 and $r$ = 350 $\mu$m. This induces an order of magnitude difference in the inverse bremsstrahlung absorption ($K_{BI} \propto n_e^2$) and thus on the amplitude and velocity of each HSW, leading to variations on the time of collision as well as on the position, density, width and shape of the final plasma slab encountered by the ps-pulse.

Despite these fluctuations, all the improvements observed in proton spectra are in good agreement with laser-plasma conditions becoming closer from the main criteria for efficient CSA \cite{Fiuza_2012,Fiuza_2013} which are i) a near-critical plasma density to get an efficient laser absorption of the ps-pulse and so a strong plasma heating (MeV), ii) a narrow plasma length to favor uniform heating, leading to a uniform shock velocity and thus the production of monoenergetic protons, iii) an exponentially decreasing output density gradient to get a uniform sheath field which preserves the monoenergetic distribution as protons are reflected by the shock.


\begin{table}
	\includegraphics[width=\columnwidth]{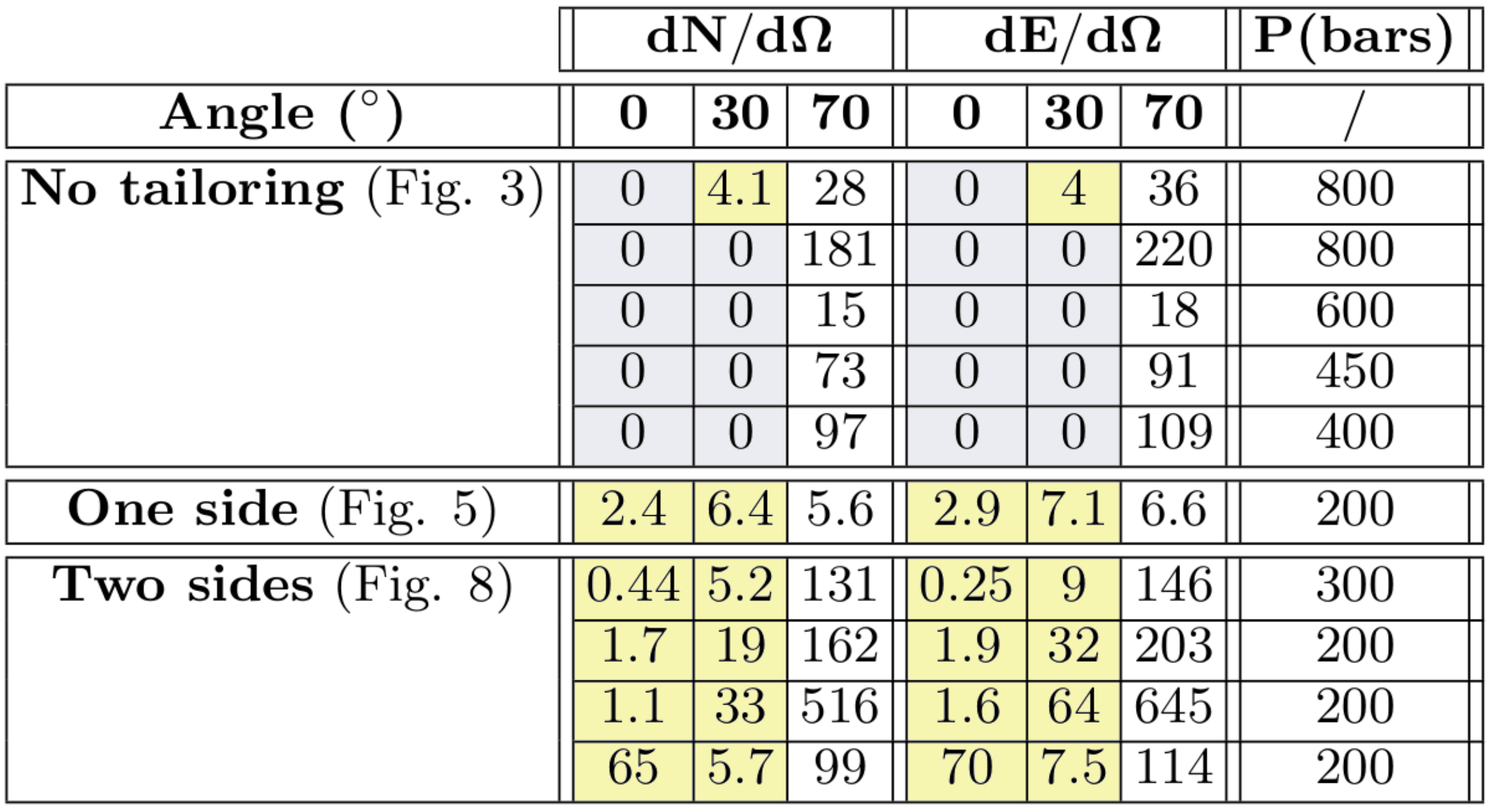}
	\caption{\textit{Number of protons and total energy of the spectra for $E_k > $ 1 MeV, measured at 0$^\circ$, 30$^\circ$ or 70$^\circ$, and for the different configurations of plasma tailoring. Obtained by integrating the spectra of Fig. \ref{Protons_wo_shaping}, \ref{Protons_shaping_face_avant} and \ref{Protons_shaping_deux_faces} between $E_k$ = 1 and 3.5 MeV. These values are expressed in $10^8$ sr$^{-1}$ for $dN/d\Omega$, and $10^8$ MeV sr$^{-1}$ for $dE/d\Omega$. Acceleration in the forward direction is highlighted with colored cells.}}
	\label{Table_proton_numbers}
\end{table}

The proton energy expected from CSA \cite{Fiuza_2012,Fiuza_2013} is of the order of $E_k$[MeV] $\sim 2 M^2 T_e$[MeV], where $M$ is the Mach number of the collisionless wave, $T_e$[MeV] $\sim  0.078 \frac {\eta} {\alpha}  a_0 \tau$(ps)$/L$(mm), $\tau$ the duration of the laser pulse, $\eta$ its absorption efficiency, and $\alpha$ is 3/2 for non-relativistic plasmas ($T_e \ll m_ec^2$) and 3 in the relativistic case. The absorption efficiency depends on $a_0$ and $n_e$. In our conditions, it is expected \cite{Tsung,Wilks1992,Fiuza_2013,Bonvalet} to be between 0.25 and 0.5. Taking $\alpha$ = 2.2 and assuming 0.05 $<L$(mm)$<$ 0.1 (smaller than the shadowgraphy size, see following section), $T_e$ is expected to be of the order of 0.35 to 1.4 MeV on both the LULI and the GSI experiments (same $a_0 \tau$). Taking for $M$ the minimum value required for the shock to reflect protons \cite{Sagdeev,Fiuza_2013}, $M = M_{cr} \sim $1.6 leads to expected proton energies of 1.8 $<E_k$(MeV) $<$ 7.2, in good agreement with our measurements.

As discussed in section \ref{Intro}, the maximum proton energy from CSA driven by a $\lambda_0$ = 1 $\mu$m laser is expected \cite{Fiuza_2012,Fiuza_2013} to occur for a plasma length close to $L_{opt} \sim$ 40 $\mu$m. The width of the plasma observed in figures \ref{ombro_LULI_shaping_deux_faces},  \ref{ombro_LULI_shaping_deux_faces_deux_temps} and \ref{ombro_spectres_GSI}-a) appears to be $\sim$ 200 to 300 $\mu$m. However, the width of the shadowgraphy could significantly overestimate those of the density profile. For example, the gas jet FWHM is $\sim$ 270 $\mu$m, while on figures \ref{ombro_LULI_GSI_sans_shaping} and \ref{ombro_streak_sans_shaping} the full width of the shadowgraphy is $\sim$ 600 to 800 $\mu$m, representing the very low wing of the density profile, 0.08 to 0.025 of the density at the jet center. The  plasma slab at the collision of the HSWs can thus be much narrower than its apparent size on the shadowgraphy.

\section{Plasma tailoring - comparison with numerical simulations}\label{hydro_3D_simulations}
\begin{figure}
	\includegraphics[width=\columnwidth]{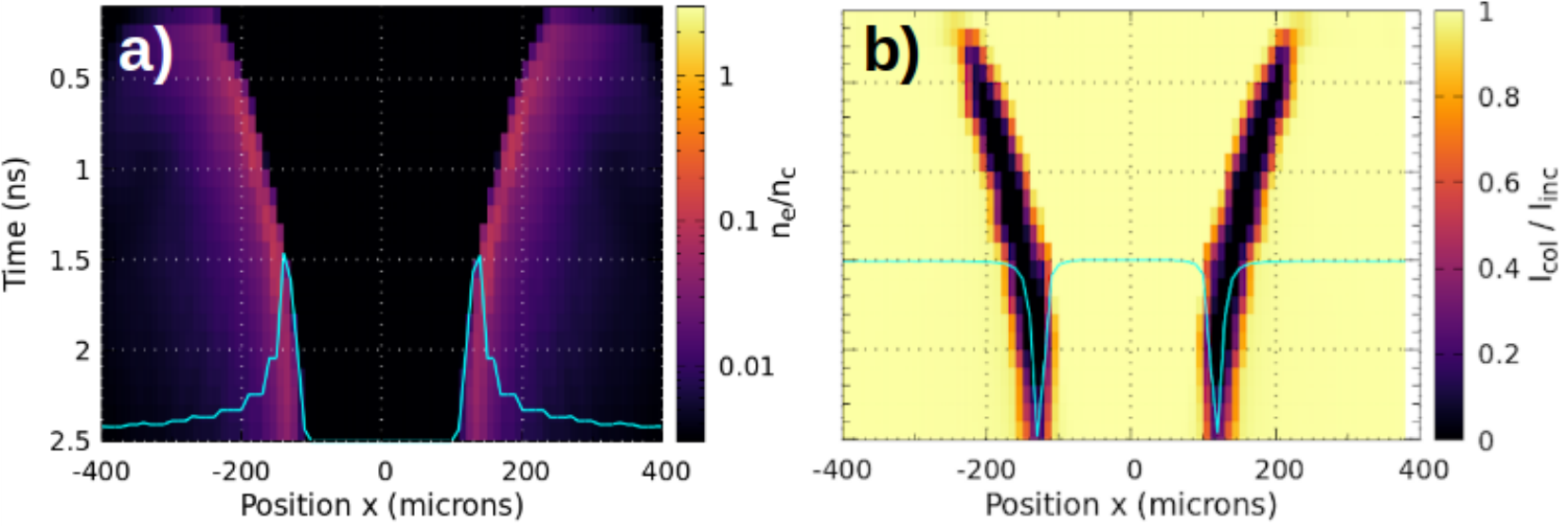}
	\caption{\label{Simul_TROLL_densite_vs_ombro_nofoot}\textit{a): Temporal evolution of the density profile along the ps-beam axis from 3D TROLL simulations. The two ns-beams are focused 300 $\mu$m from both sides of the gas jet center. The initial maximum plasma density (at $x$ = 0) is $n_e^0 = 2\times 10^{20}$ cm$^{-3}$ ($n_c/5$). The spatial intensity profile of the ns-beams is $I(r) = 5.4\times 10^{14} e^{-(r/25)^2}$, and the energy in each beam is 6 J. b) Post-processing of the shadowgraphy diagnostic from the TROLL outputs. The cyan line-outs are density and shadowgraphy profiles at $t$ = 2.5 ns.}}
\end{figure}

\begin{figure}
	\includegraphics[width=\columnwidth]{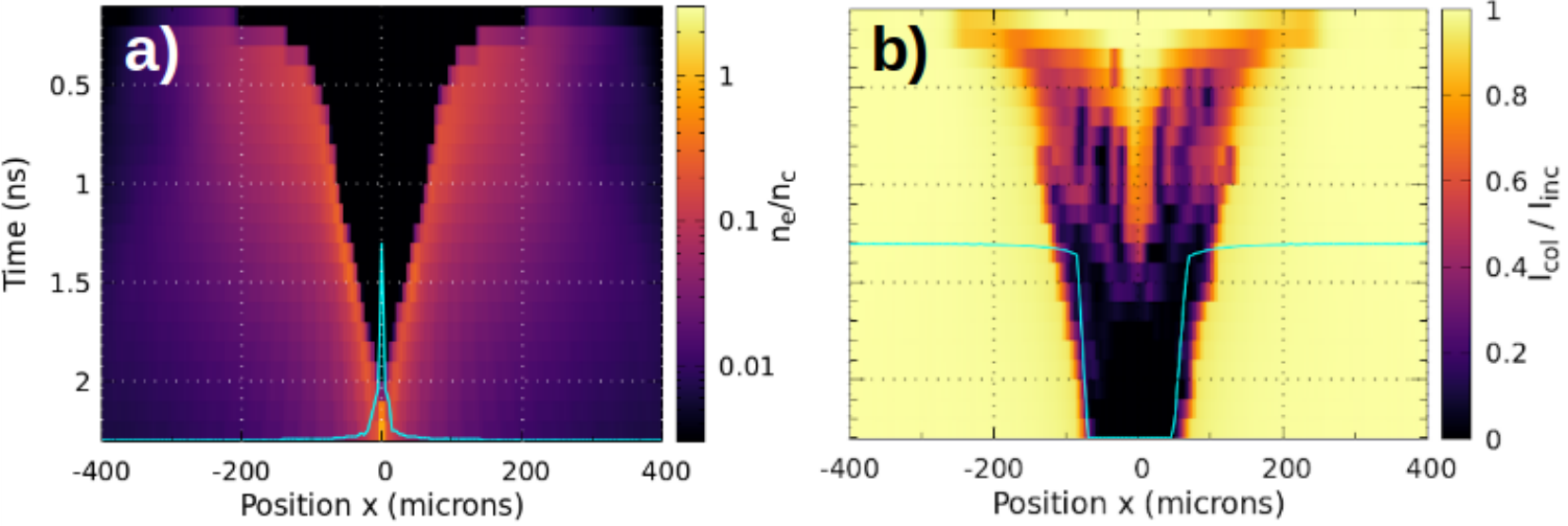}
	\caption{\label{Simul_TROLL_densite_vs_ombro_wfoot}\textit{Same as Fig. \ref{Simul_TROLL_densite_vs_ombro_nofoot} but for a spatial profile of the ns-beams with a low-intensity wing: $I(r) = 3\times 10^{14} [e^{-(r/25)^2} + 0.0172 e^{-(r/110)^2}]$ W/cm$^2$, and 4.5 J in each beam. The cyan line-outs are density and shadowgraphy profiles at $t$ = 2.3 ns.}}
\end{figure}

\begin{figure}
	\includegraphics[width=\columnwidth]{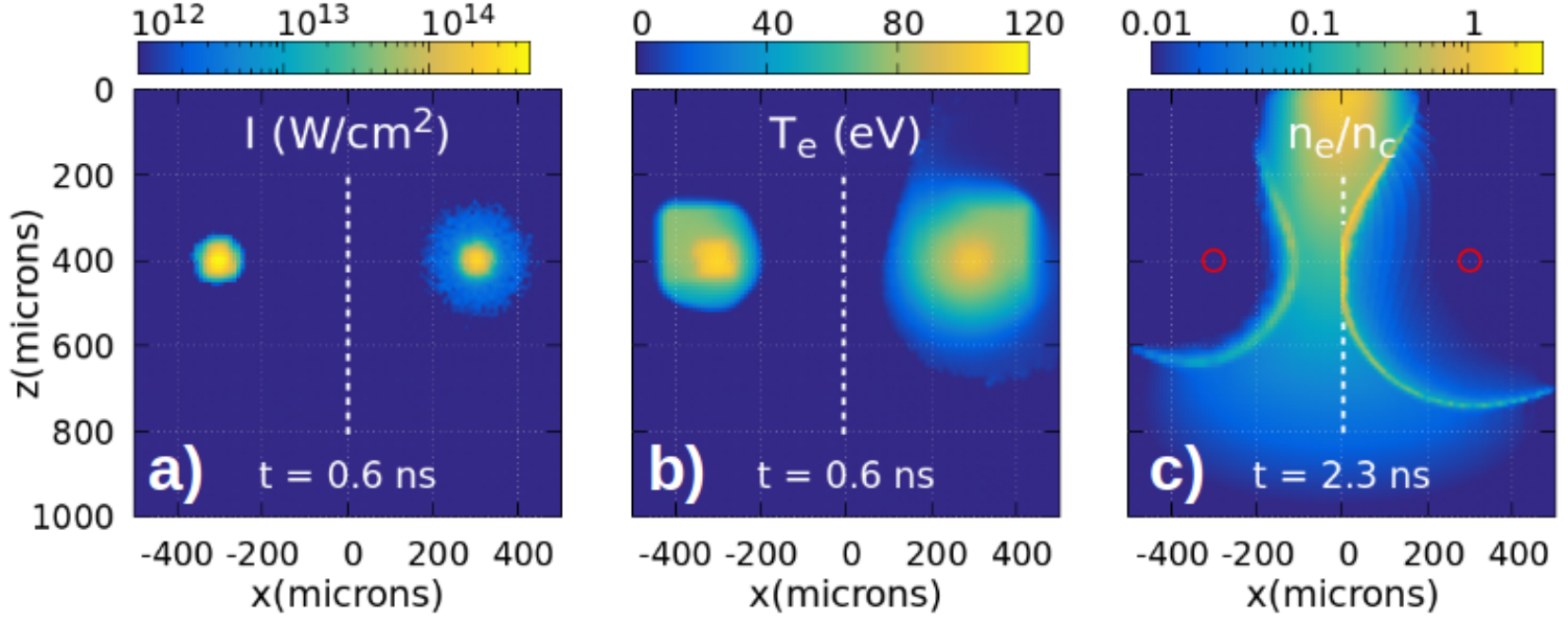}
	\caption{\label{Simul_TROLL_profils_nofoot_wfoot}\textit{Spatial profiles of a) the laser focal spot, b) the electron temperature, c) the electron density for the case without the wing in the intensity profile (left, Fig. \ref{Simul_TROLL_densite_vs_ombro_nofoot}) or with (right, Fig. \ref{Simul_TROLL_densite_vs_ombro_wfoot}). a) et b) are at 0.6 ns, the maximum of the laser, c) is at 2.3 ns (time at which the HSWs collide in the case with the wing).}}
\end{figure}

The CSA strongly depends on the plasma length $L$. Our previous numerical study \cite{Marques} indicated that the width of the density profile at the collision of the two HSWs was below or of the order of tens of microns while, as previously said, the dark part of the shadowgraphy images in Fig. \ref{ombro_LULI_shaping_deux_faces} and \ref{ombro_spectres_GSI}-a) has a width of the order of 200-300 $\mu$m. This dark part originates from light rays that are so refracted by the density gradient that they are not collected by the lens of the imaging system. This width thus depends on the aperture of the diagnostic (3.5$^{\circ}$ on the LULI experiment, 2$^{\circ}$ on GSI) and can be significantly larger than the FWHM of the density profile. To estimate this difference we performed 3D hydrodynamics simulations, and post-processed the probe beam propagation (refraction and inverse bremsstrahlung absorption) to simulate the shadowgraphy diagnostic. As in our previous study \cite{Marques}, we used the radiation-hydrodynamics code TROLL \cite{Lefebvre}, but in a three-dimensional geometry, allowing to use as inputs the experimentally measured 3D density profile of the gas jet. We also used the experimental temporal profile of the ns-beams. Figure \ref{Simul_TROLL_densite_vs_ombro_nofoot}-a) shows the temporal evolution of the density profile along the ps-beam axis. As already detailed in Ref. \cite{Marques}, each of the two ns-beams, focused 300 $\mu$m from both sides of the gas jet center ($x=0$) generates a HSW. Only the part of the cylindrical HSW that converges towards the gas jet center leads to a high density jump, while the part traveling towards vacuum is not visible because of its quick spread and decrease in amplitude. The initial maximum plasma density (at $x$ = 0) is $n_e^0 = 2\times 10^{20}$ cm$^{-3}$ ($n_c/5$), as for the shot presented in Fig. \ref{ombro_LULI_shaping_deux_faces}-a). The spatial intensity profile of the ns-beams is a Gaussian of 25 $\mu$m radius. Fig. \ref{Simul_TROLL_densite_vs_ombro_nofoot}-b) is the simulation of the shadowgraphy diagnostic. Compared to Fig. \ref{ombro_LULI_shaping_deux_faces}-a), after 2.5 ns the HSWs still have not reached the gas jet center. Also, at the HSW creation, the experimental shadowgraphy has a large "horizontal" shape while the simulated one is much more localized (narrow). In the experiment, the laser intensity profile was in fact composed not only of the narrow high intensity part, but also of a lower intensity wing: $I(r) = 3\times 10^{14} [e^{-(r/25)^2} + 0.0172 e^{-(r/110)^2}]$ W/cm$^2$. Despite its lower intensity, this large wing contained $\sim$ 25 $\%$ of the laser energy, enough to contribute to the HSW excitation. Fig. \ref{Simul_TROLL_densite_vs_ombro_wfoot} shows the evolution of the density profile and of the associated shadowgraphy when the wing of the ns-beams is taken into account, and for the same laser energy (4.5 J in each beam) as the shot in Fig. \ref{ombro_LULI_shaping_deux_faces}. The simulated shadowgraphy is very similar to the experimental one (Fig. \ref{ombro_LULI_shaping_deux_faces}-a)), reproducing the initial "horizontal" part followed by two large traces that converges when the two HSWs collide after $\sim$ 2.3 ns, at the jet center. The effect of the wing of the focal spot on the HSW evolution is also illustrated in Fig. \ref{Simul_TROLL_profils_nofoot_wfoot}, that compares the profiles of a) the laser focal spot, b) the electron temperature, c) the electron density for the case without (left) or with (right) the wing. Even if the intensity is higher without the wing, because the energy deposition is more localized, the HSWs start with a smaller diameter and further from the jet center, avoiding them to collide. With the wing, the HSWs are generated with a larger diameter, enabling them to reach faster the jet center. A point to underline is that the density digging downstream the HSW seems nevertheless more efficient without wing (more localized energy deposition), as can also be observed comparing Fig. \ref{Simul_TROLL_densite_vs_ombro_nofoot}-a) and \ref{Simul_TROLL_densite_vs_ombro_wfoot}a). Fig. \ref{Simul_TROLL_profils_nofoot_wfoot}-c) also shows that the simulations reproduce very well the "croissant-like" shape observed experimentally (Fig. \ref{ombro_LULI_shaping_face_avant}, \ref{ombro_LULI_shaping_deux_faces}, \ref{ombro_LULI_shaping_deux_faces_deux_temps} and \ref{ombro_spectres_GSI}).  The influence of the wing is also observed in Fig. \ref{ombro_spectres_GSI}-a) where the square shape of the GSI ns-beams is retrieved in the shape of the HSWs.

Let us note that, since hydrogen is very quickly ionized, a full ionization at the beginning of the simulation was imposed and, to avoid an early gas jet expansion, an initial temperature of 300 K was imposed. This has no effect on the plasma evolution induced by the ns-beams. However, in the regions not affected by these beams or by the HSWs they produce, where the medium should still be in its gas state, its refraction index as well as the inverse bremsstrahlung absorption of the low-intensity probe beam are strongly overestimated. To avoid this artifact, in the post-processing of the shadowgraphy diagnostics the ionization state was corrected using the Saha ionization equation. The density map presented in Fig. \ref{Simul_TROLL_densite_vs_ombro_nofoot}-a) includes this correction. Without this correction, the shadowgraphy at the early times is completely dark at the jet center, while the gas is still not ionized and should fully transmit the probe beam (as experimentally observed and mentioned in section \ref{1beam-tailoring} and \ref{2beam-tailoring}).

The normalized line-outs in Fig \ref{Simul_TROLL_densite_vs_ombro_wfoot}-a) and -b) show respectively the density and shadowgraphy profiles at the collision time ($t=2.3$ ns). The width of the shadowgraphy is $\sim$ 150 $\mu$m, much larger than the width of the plasma slab, $\lesssim$ 20 $\mu$m. Applying this same factor difference on the experimental shadowgraphies would indicate that the plasma slab at the collision time had a thickness of the order of 30 to 40 $\mu$m, similar to the predicted optimal value $L_{opt}$ for CSA (section \ref{Intro}).

From Fig. \ref{Simul_TROLL_densite_vs_ombro_wfoot}-a), the peak density at the collision time is $\sim 3 n_c$. By conservation of the number of particles, if the experimental plasma slab is of the order or more than 2 times thicker than in the simulations, the peak density should be lower by this same factor, of the order or less than 1.5 $n_c$.

\section{Discussion}\label{Discussion}
Even if the thickness of the plasma slab is close to optimum, its density needs also to be adjusted to an optimum value which depends on the laser intensity. To drive a strong collisionless shock and accelerate ions at high energies, a high laser absorption is required (see section \ref{2beam-tailoring}), which implies $n_e \sim \gamma n_c$. With $a_0\sim$ 4 at LULI and 8 at GSI, this leads to $n_e/n_c \sim$ 3 and 5.7 respectively. Our previous numerical study \cite{Bonvalet} also showed that the maximum proton energy drops quickly below this optimum value. The density of the plasma slab was thus probably too low for an optimum proton acceleration. An easy way to produce plasma slabs of higher density is to increase the backing pressure of the gas jet, as long as the residual density in the wing of the slab does not disturb the propagation of the ps laser beam.

The efficiency of the acceleration depends on the intensity of the ps-laser, but also on the size of its focal spot \cite{Fiuza_2013}. The shock width (which is close to the laser spot size $W_0$) has to be large enough such that the plasma, expanding transversely at the sound speed, does not leave the shock region before the acceleration occurs. Assuming an isothermal expansion, this condition yields $W_0 \gtrsim L/M$. Taking for $M$ the minimum value required for the shock to reflect protons, $M_{cr} =$ 1.6, and $L = $ 40 $\mu$m leads to $W_0 \gtrsim$ 25 $\mu$m. The focal spot on the LULI and GSI experiments (FWHM $\sim$ 10 $\mu$m) was thus probably too small for an optimum acceleration. With a larger focal spot the shape of the shock will be more flat, favoring forward acceleration, at higher energy. The larger surface might also increase the number of accelerated protons.

The shot-to-shot fluctuations of the ns-beams, in terms of position, spatial distribution and intensity, generate spatial and temporal fluctuations on the HSWs, and thus on the resulting plasma slab. Coupled to the temporal jitter between the ns and ps pulses (between 100 and 200 ps in our experiment), significant differences in laser-plasma coupling and heating can be induced, which is thought to be at the origin of shot-to-shot fluctuations of the proton beam. It also indicates that the condition for efficient CSA are still not fully reached. Hosing instability\cite{Ceurvorst} could also modify the direction of propagation of the ps-pulse, ultimately leading to fluctuations in the direction of the collisionless shock and of the proton beam. This affects more significantly the detection in the forward direction, mainly sensitive to CSA.

The clearly peaked spectra observed with the two-sides tailoring indicate that even if the plasma slab is relatively narrow, the density gradient at its backside is not very sharp, avoiding TNSA, and present a "slowly" decreasing profile with a small and constant sheath electric field in favor of monoenergetic acceleration \cite{Fiuza_2012}.

\section{Conclusions}
We recently proposed a new technique of plasma tailoring by laser-driven hydrodynamic shockwaves generated on both sides of a gas jet. In the continuation of this numerical work, we studied experimentally the influence of the tailoring on proton acceleration driven by a high-intensity ps-laser, in three cases: without tailoring, by tailoring only the entrance side of the ps-laser, or both sides of the gas jet. Without tailoring the acceleration is transverse to the laser axis, with a low energy exponential spectrum, produced by Coulomb explosion. When the front side of the gas jet is tailored, a forward acceleration appears, that is significantly enhanced when both the front and back sides of the plasma are tailored. This forward acceleration produces higher energy protons (up to 5.5 MeV), with a peaked spectrum, and is in good agreement with the mechanism of Collisionless Shock Acceleration. The total number of accelerated protons is also enhanced by this two-sides tailoring. The spatio-temporal evolution of the plasma profile was characterized by optical shadowgraphy of a probe beam. The refraction and absorption of this beam was simulated by post-processing 3D hydrodynamic simulations of the plasma tailoring. The dynamics of the hydrodynamic shockwaves was well reproduced only if the low-intensity wing in the focal spot of the tailoring beams was taken into account. Comparison of the simulated shadowgraphy with the experimental ones allowed to estimate the thickness ($\sim$ 30-40 $\mu$m) and density ($\sim$ 1.5 $n_c$) of the plasma slab produced by tailoring both sides of the gas jet. These values are close to those required to trigger the CSA. However, the shot-to-shot fluctuations of the forward proton beam indicate that the set of criteria for an optimum CSA was still not reached. Improving the focal spot quality of the tailoring ns-beams, increasing the backing pressure of the gas jet as well as the focal spot size of the ps-beam should enable to reach a more stable CSA, generating a proton beam with a larger number of protons, at higher energy and of lower spectral width and emittance.

\begin{acknowledgments}
The authors would like to thank the LULI staff and the GSI-PHELIX staff for their contribution. This work has received funding from the Fédération de recherche PLAS@PAR. Results presented here are partially based on the experiment P189, which was performed at the PHELIX infrastructure at GSI Helmholtzzentrum fuer Schwerionenforschung, Darmstadt (Germany) in the context of FAIR Phase-0. The research leading to the PHELIX-GSI results has received funding from the European Union’s Horizon 2020 research and innovation program under Grant Agreement No. 871124 Laserlab-Europe, and by Grant ANR-17-CE30- 0026-Pinnacle from Agence Nationale de la Recherche.
\end{acknowledgments}

\section*{Conflict of Interest}
The authors have no conflicts to disclose.

\section*{data availability}
The data that support the findings of this study are available from the corresponding author upon reasonable request.

\bibliography{Proton_CSA_in_laser_tailored_plasma}

\end{document}